# Steganography Using Adaptive Pixel Value Differencing(APVD) of Gray Images Through Exclusion of Overflow/Underflow

J. K. Mandal[1] and Debashis Das[2]

[1] Department of Computer Science and Engineering,
University of Kalyani, Kalyani,
West Bengal, India
`jkm.cse@gmail.com`

[2] Department of Computer Science and Engineering,
University of Kalyani, Kalyani,
West Bengal, India
debashisitnsec@gmail.com

**ABSTRACT**

In a gray scale image the pixel value ranges from 0 to 255. But when we use pixel-value differencing (pvd) method as image steganographic scheme, the pixel values in the stego-image may exceed gray scale range. An adaptive steganography based on modified pixel-value differencing through management of pixel values within the range of gray scale has been proposed in this paper. PVD method is used and check whether the pixel value exceeds the range on embedding. Positions where the pixel exceeds boundary has been marked and a delicate handle is used to keep the value within the range. From the experimental it is seen that the results obtained in proposed method provides with identical payload and visual fidelity of stego-image compared to the pvd method.

## 1. INTRODUCTION

Now-a-days Internet has become the most popular communication media for message transmission in different places of the world. Two schemes are used to protect secret messages from being captured during transmission. One is encryption where the secret information is encoded in another form by using a secret key before sending, which can only be decoded with secret keys. The most popular encryption techniques are DES, RSA etc. Other way is steganography which is a technique of hiding secret information into a cover media or carrier. If the cover media is a digital image, it is called cover image and the cover image with hidden data is called stego-image. Steganographic technique can be used in military, commercial, anti-criminal and so on. There are various steganographic techniques available where a digital image is used as a carrier. The most common and simplest method is least-significant-bit (LSB) substitution, where the LSB position of each pixel of the cover image is replaced by one bit of secret data. Wang et al.[6] proposed a method to embed data by using genetic algorithm to improve the quality of the stego-image. However, genetic algorithm takes more computational time. Chang et al. Proposed[8] an efficient dynamic programming strategy to reduce the computational time. Chan and Cheng [10] proposed to embed data by simple LSB substitution with an optimal pixel adjustment process. Wu and Tsai[1] proposed a new scheme to hide more data with outstanding quality of stego-image pixel-value-differencing (PVD) method. Thereafter, based on PVD method various approaches have been proposed [2,3,4,7,9]. In PVD the range may exceed -64 to 319 in embedding. In this paper, a range managed steganographic approach, using PVD has been proposed. In APVD method a gray scale digital image has been used as a cover image where pixel values ranged between 0 and 255 and the pixel values of stego-image will not exceed the gray scale range. The proposed method will provide the same hiding capacity as of the original PVD method with acceptable stego image quality .

## 2. REVIEW OF PVD METHOD

In PVD method[1], gray scale image is used as a cover image with a long bit-stream as the secret data. At first the cover image is partitioned into non-overlapping blocks of two consecutive pixels, $p_i$ and $p_{i+1}$. From each block the difference value $d_i$ is calculated by subtracting $p_i$ from $p_{i+1}$. The set of all difference values may range from -255 to 255. Therefore, $|d_i|$ ranges from 0 to 255. The blocks with small difference value locates in smooth area where block with large difference values are the sharp edged area. According to the properties of human vision, eyes can tolerate more changes in sharp-edge area than smooth area. So, more data can be embedded into edge area than smooth areas. Therefore, in PVD method a range table has been designed with n contiguous ranges $R_k$ (where k=1,2,…,n) where the range is 0 to 255. The lower and the upper bound are denoted as $l_k$ and $u_k$ respectively, then $R_k \in [l_k, u_k]$. The width of $R_k$ is calculated as $w_k = u_k - l_k + 1$. $w_k$ decides how many bits can be hidden in a pixel block. For security purpose $R_k$ is kept as a variable, as a result, original range table is required to extract the embedded data. The embedding algorithm is given as algorithm 1

**Algorithm 1**:
1. Calculate the difference value $d_i$ of two consecutive pixels $p_i$ and $p_{i+1}$ for each block in the cover image. This is given by $d_i = |p_{i+1} - p_i|$.
2. Compute the optimal range where the difference lies in the range table by using $d_i$. This is calculated as $R_i = \min(u_k - d_i)$, where $u_k \geq d_i$ for all $1 \leq k \leq n$
3. Compute the number of bits 't' to be hidden in a pixel block can be defined as $t = \lfloor \log_2 w_i \rfloor$. where $w_i$ is the width of the range in which the pixel difference $d_i$ is belonging
4. Read t bits from binary secret data and convert it into its corresponding decimal value b. For instance if t=010, then b=2
5. Calculate the new difference value $d_i'$ which is given by $d_i' = l_i + b$
6. Modify the values of $p_i$ and $p_{i+1}$ by the following formula:

$$(p_i', p_{i+1}') = \begin{cases} (p_i + \lceil m/2 \rceil, p_{i+1} - \lfloor m/2 \rfloor), & \text{if } p_i \geq p_{i+1} \text{ and } d_i' > d_i \\ (p_i - \lfloor m/2 \rfloor, p_{i+1} + \lceil m/2 \rceil), & \text{if } p_i < p_{i+1} \text{ and } d_i' > d_i \\ (p_i - \lceil m/2 \rceil, p_{i+1} + \lfloor m/2 \rfloor), & \text{if } p_i \geq p_{i+1} \text{ and } d_i' \leq d_i \\ (p_i + \lceil m/2 \rceil, p_{i+1} - \lfloor m/2 \rfloor), & \text{if } p_i < p_{i+1} \text{ and } d_i' \leq d_i \end{cases}$$

where $m = |d_i' - d_i|$. Now we obtain the pixel pair $(p_i', p_{i+1}')$ after embedding the secret data into pixel pair $(p_i, p_{i+1})$. Repeat step 1-6 until all secret data are embedded into the cover image. Hence we get the stego-image.

When extracting the hidden information from the stego-image, original range table is required. At first partition the stego-image into pixel blocks, containing two consecutive non-overlapping pixels each. Calculate the difference value for each block as $d_i' = |p_i' - p_{i+1}'|$. Then find the optimum range $R_i$ of $d_i'$. Then b' is obtained by subtracting $l_i$ from $d_i'$. Convert b' into its corresponding binary of 't' bits, where $t = \lfloor \log_2 w_i \rfloor$. These t bits are the hidden secret data obtained from the pixel block $(p_i', p_{i+1}')$.

## 3. PROPOSED METHOD

In PVD method pixel values in the stego image may exceed the gray scale range which is not desirable as it may leads to improper visualisation of the stego image. In this section we introduce a method to overcome this problem. In the proposed method we have used the original PVD method to embed secret data. If any pixel value exceeds the range (0 to 255), then check the bit-stream 't' to be hidden. If MSB(most significant bit) of the selected bit stream 't' is 1 then we embed one less number of bits, where MSB position is discarded from t; otherwise the bit number of hidden data depends on $w_i$. For instance, if pixel value exceeds the range and selected bit-stream t=101, then set t=01 and embed it. If it is seen that the pixel value again exceeding range, then embed the value at one pixel, rather than both pixels(of the pixel block), which will not exceed the range after embedding; where the other pixel is kept unchanged. It will keep the pixel values within the range because both pixels of a block cannot exceed at the same time as per the PVD method by Wu and Tsai. Keep the information within each block, whether one less bit is embedded or not, as overhead. The embedding algorithm is presented in section 3.1. Figure 1 shows the block diagram of the embedding algorithm.

### 3.1. Embedding Algorithm

Step 1: Calculate the difference value di for each block of two consecutive non-overlapping pixels $p_i$ and $p_{i+1}$, is given by $d_i=|p_i-p_{i+1}|$.

Step 2: Find optimal range $R_i$ for the $d_i$ such that $R_i =\min(u_i-d_i)$, where $u_i \geq d_i$. Then $R_i \in [l_i, u_i]$ is the optimum range where the difference lies.

Step 3: Compute the amount of secret data bits t from the width $w_i$ of the optimum range, can be defined as $t= \lfloor \log 2\ w_i \rfloor$.

Step 4: Read t bits and convert it into a decimal value b. Then calculate the new difference value by the formula $d_i'=l_i+b$.

Step 5: Now, calculate the pixel values after embedding t bits ($p_i'$, $p_{i+1}'$) by original PVD method.

Step 6: Check the embedded pixel values whether it exceed the gray-level range or not. If it exceeds then check the embedded bit-stream t. Otherwise go to step 7.

Step 6.1: If the left most position of the bit-stream is 1 then select 't' by discarding one bit from its left most position. Convert 't' bits into its corresponding decimal value b and find new difference value as $d_i'=l_i+b$. Otherwise do not discard any bit from 't' and calculate $d_i'$.

Step 6.2: Calculate new pixel values ($p_i'$, $p_{i+1}'$) using original PVD method and check again if it is in the gray range. If it is in the range then go to step 7. Otherwise do the following:

$(p_i', p_{i+1}')=(p_i-m, p_{i+1})$,    if $p_{i+1} \geq p_i$ and $p_{i+1}$ crossing the upper range(i.e 255);
$(p_i', p_{i+1}')=(p_i, p_{i+1}-m)$,    if $p_{i+1}<p_i$ and $p_i$ crossing the upper range(i.e 255);
$(p_i', p_{i+1}')=(p_i, p_{i+1}+m)$,    if $p_{i+1} \geq p_i$ and $p_i$ crossing the lower range(i.e 0);
$(p_i', p_{i+1}')=(p_i+m, p_{i+1})$,    if $p_{i+1}<p_i$ and $p_{i+1}$ crossing the lower range(i.e 0).

where $m=|d_i'-d_i|$.

Step 7: Now, the pixel block ($p_i$, $p_{i+1}$) is replaced by ($p_i'$, $p_{i+1}'$).

Step 8: To keep the information whether 't' bits or 't-1' bits has been embedded, do the following for each modified block :

Step 8.1:
If no bit has been discarded then do the following:

| | LSB of $P'_i$ | LSB of $P'_{i+1}$ | then do |
|---|---|---|---|
| i) | 0 | 0 | $P'_{i+1}+1$ |
| ii) | 0 | 1 | |

    a) $P'_{i+1}<255$ and $P'_I \geq 0$     $P'_{i+1}+1$
    b) $P'_i >0$ and $P'_{i+1} =255$     $P'_i-2$ and $P'_{i+1} -1$
    c) $P'_i=0$, $P'_{i+1}=255$     $P'_{i+1}$

| iii) | 1 | 0 | $P'_i$ -1 |
| iv) | 1 | 1 | $P'_i$ -1 |

**Step 8.2:**
    One bit has been discarded

| | LSB of $P'_i$ | LSB of $P'_{i+1}$ | then do |
|---|---|---|---|
| i) | 0 | 0 | $P'_i$ +1 |
| ii) | 0 | 1 | $P'_i$ +1 |
| iii) | 1 | 0 | |
| | a) $P'_{i+1} > 0$ and $P'_i \leq 255$ | | $P'_{i+1}$ -1 |
| | b) $P'_i < 255$ and $P'_{i+1} = 0$ | | $P'_i$+2 and $P'_{i+1}$+1 |
| iv) | 1 | 1 | $P'_{i+1}$ -1 |

**Step 9:** Now we get the stego blocks and hence the stego image.

## 3.2. Extraction Algorithm

Figure 1 shows the block diagram of the extraction algorithm. The steps used for extracting the hidden data are as follows :

**Step 1:** Partition the stego-image into pixel blocks consist of two consecutive non-overlapping pixels.

**Step 2:** Check the LSB positions of Pi for each block and do the following :

| LSB of $P_i$ | Convert it as |
|---|---|
| 0 | $P_i$ +1 |
| 1 | $P_i$ -1 |

**Step 3:** Now calculate the difference value di of two consecutive pixels of each block by using the formula $d_i = |p_i - p_{i+1}|$.

**Step 4:** Find the appropriate range $R_i$ for the difference $d_i$.

**Step 5:** Extract 't' bits, by the extracting method of original PVD. where $t = \lfloor \log_2 w_i \rfloor$, wi is the width of the optimal range $R_i$.

**Step 6:** Check the LSB of $P_i$. If it is 1, replace the MSB position of extracted 't' bits with '1'. Otherwise do nothing.

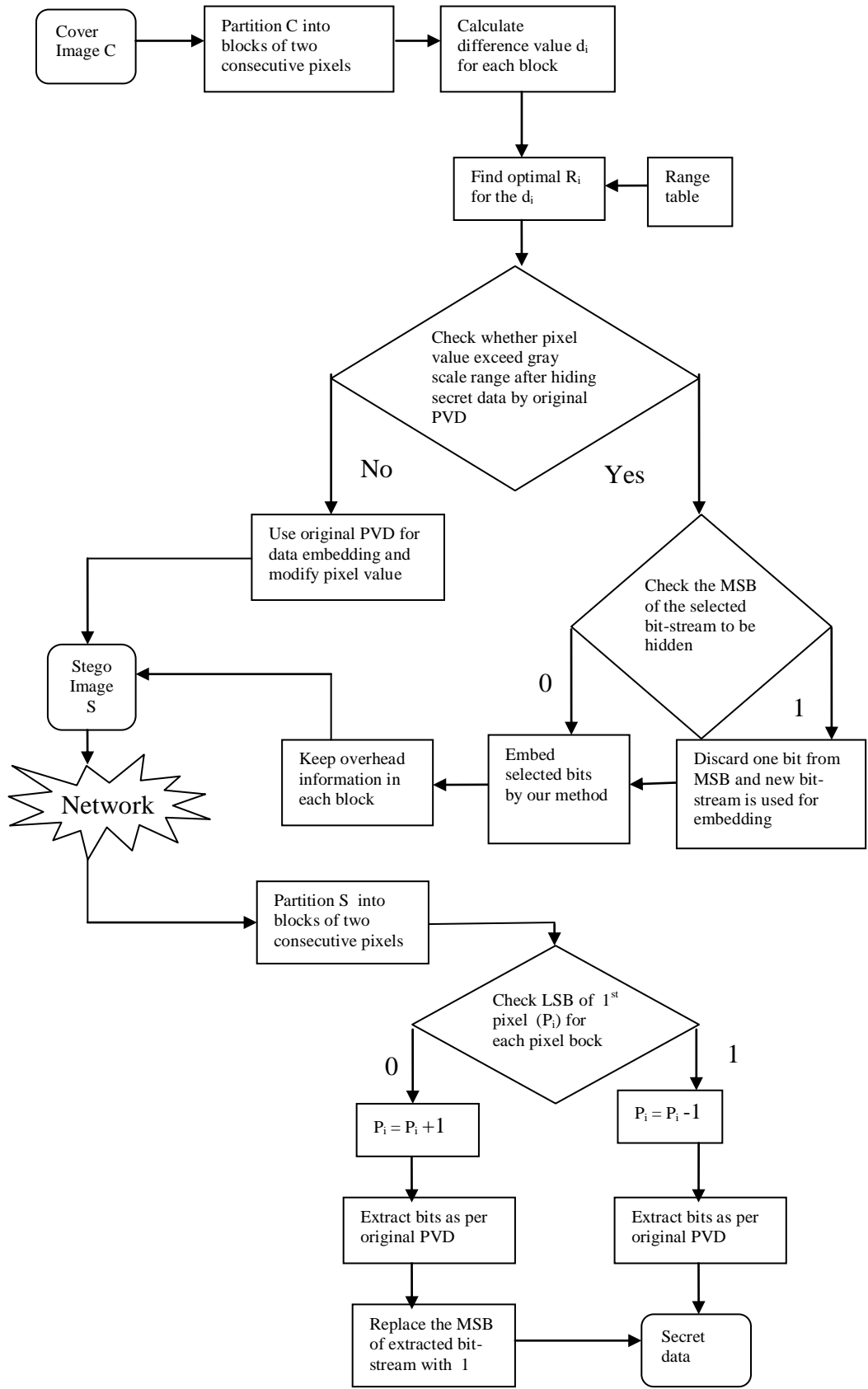

Fig. 1 *Block diagram of embedding and extraction algorithm*

## 4. EXPERIMENTAL RESULTS

The problem of overshooting gray-level range in PVD has been removed which results no effect on hiding capacity. we have used C programming language to implement our scheme and Wu-Tsai's PVD scheme. The range table width used here are $w_i$ ={ 8, 8, 16, 32, 64, 128 }.We have used cover images of size 512*512 and hide a digital image as the secret information. We have used the Peak-Signal-to-Noise ratio (PSNR) to evaluate the quality of stego-image. The experimental results are shown in Fig. 3 and Fig. 4. The comparison, of message capacity and PSNR value, between proposed method and original PVD method is shown in Table 2. The PSNR value and capacity of each stego-image is given as average value by executing 10 rounds using standard digital images where the hidden information are different.

## 5. Analyses and Discussion

From the experimental results we can see that PSNR values are changing between -0.62 to +0.32 dB and capacity remains same compared to original PVD method. According to the proposed method, if pixel value exceeds gray scale range, one bit is discarded from the selected bit-stream to be hidden. So, decimal value of the reduced bits will be half or less than half. As a result the distortion of the pixel value in the stego-image will be less. On the other hand, for keeping the overhead information we are adding or subtracting some values with the pixel values. This can increase the distortion of the image. It should be noted that whether the pixel exceeds range or not, a track is kept as information in every pixel block. So, for the combined effect PSNR value somewhere increases and also decreases in some cases but of a little bit.

Steganographic technique of C.M. Wang et al. [11] devised for increased PSNR of the stego image with pixel value differencing and modulus function. They have also removed the problem of range overflow in their paper. But the data embedding procedure was completely different from original PVD devised by Wu and Tsai. So, it can be said that the original PVD was still carrying the range overflow problem. Therefore we eliminate the problem of original PVD in our proposed method where as, a completely different technique has been used from the method proposed by C.M. Wang et al.

In the method of C.M. Wang et al. [11], they have used the pixel value differences to find the number of embedding bits for each pixel block. Then a modulus function has been applied to calculate the remainder values as follows;

$$P_{rem(i,x)} = P_{(i,x)} \bmod t_i'$$
$$P_{rem(i,y)} = P_{(i,y)} \bmod t_i \quad \text{[where } t_i' \text{ is the decimal value of secret bits } t_i\text{]}$$
$$F_{rem(i)} = (P_{(i,x)} + P_{(i,y)}) \bmod t_i'$$

Secret bit stream is embedded by altering two pixels such that $F_{rem(i)} = t_i'$. There may be eight different cases for embedding bits to achieve minimum distortion [11].

After the embedding process is over, handling the falling-off boundary problem is done as follows:

(P''$_{(i,x)}$, P''$_{(i,y)}$) = ( P'$_{(i,x)}$ − (2$^{ti}$)/2, P'$_{(i,y)}$ − (2$^{ti}$)/2 ) ;  if  P'$_{(i,x)}$>255 or P'$_{(i,y)}$>255  and d$_i$<128
(P''$_{(i,x)}$, P''$_{(i,y)}$) = ( P'$_{(i,x)}$ + (2$^{ti}$)/2, P'$_{(i,y)}$ + (2$^{ti}$)/2 ) ;  if  P'$_{(i,x)}$<0 or P'$_{(i,y)}$<0      and d$_i$<128
(P''$_{(i,x)}$, P''$_{(i,y)}$) = ( 0, P'$_{(i,y)}$ + P'$_{(i,x)}$ ) ;                if  P'$_{(i,x)}$<0 and P'$_{(i,y)}$≥ 0     and d$_i$>128
(P''$_{(i,x)}$, P''$_{(i,y)}$) = ( P'$_{(i,x)}$ + P'$_{(i,y)}$, 0 ) ;                if  P'$_{(i,x)}$ ≥ 0 and P'$_{(i,y)}$ <0    and d$_i$>128
(P''$_{(i,x)}$, P''$_{(i,y)}$) = ( 255, P'$_{(i,y)}$ +(P'$_{(i,x)}$-255)) ;        if  P'$_{(i,x)}$ >255 and P'$_{(i,y)}$≥0   and d$_i$>128
(P''$_{(i,x)}$, P''$_{(i,y)}$) = ( P'$_{(i,x)}$ +(P'$_{(i,y)}$-255), 255) ;        if  P'$_{(i,x)}$ ≥0 and P'$_{(i,y)}$ >255  and d$_i$>128

In contrast, we have used original PVD method for data embedding and problem of boundary overshooting has been managed during embedding process. If pixel value overflow, one bit is discarded from MSB position and rest of the bit-stream is embedded, as shown in embedding algorithm (section 3.1), if the problem still remains, then it has been handled using the formula given in step 6.2 of embedding algorithm. Furthermore, we have used the same method, for every difference value (d$_i$), to manage the range of pixel where C.M. Wang et al. used different techniques for d$_i$>128 and d$_i$<128. Table 1 gives a comparative statement between C.M. Wang et al. [11] and proposed methods where less distortions of pixels are occurred.

Table1: **Comparison of proposed method with C.M. Wang et al. [11]**

| Method | Input pixel block | After embedding '111' | After managing overflow | Pixel distortion |
|---|---|---|---|---|
| C.M. Wang et al.[11] | [254 , 255] | [255 , 256] | [251 , 252] | [3 , 3] |
| Proposed method | [254 , 255] | [251 , 258] | [252 , 255] | [2 , 0] |

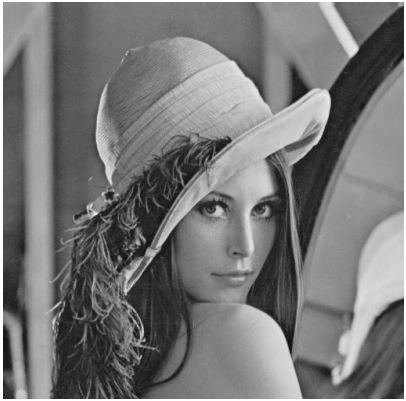 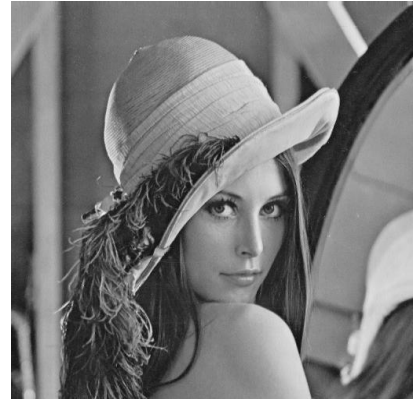

a                         b

Fig. 3  Cover image and stego-image

a. cover image Lena       b. stego-image on embedding 51370 bytes; PSNR is 40.61 dB.

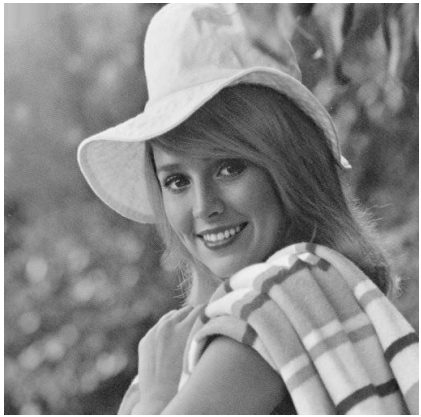 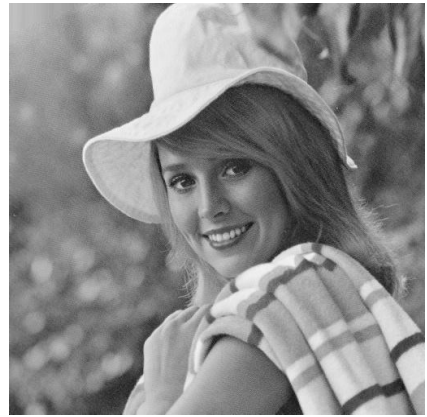

c                         d

Fig. 4  Cover image and stego-image

c. cover image Elaine       d. stego-image on embedding 51070 bytes; PSNR is 41.61 dB.

Table 2: **Comparison of results of the proposed and Wu and Tsai's method**

| Cover image (512× 512) | Wu and Tsai's method | | Our method | |
|---|---|---|---|---|
| | Capacity | PSNR | Capacity | PSNR |
| Lena | 51370 | 41.70 | 51370 | 40.61 |
| Baboon | 57583 | 36.86 | 57583 | 36.67 |
| Tank | 50495 | 42.54 | 50495 | 42.05 |
| Airplane | 49735 | 42.31 | 49735 | 42.63 |
| Truck | 50061 | 43.05 | 50061 | 42.56 |
| Elaine | 51070 | 42.09 | 51070 | 41.47 |
| Couple | 51600 | 40.33 | 51600 | 40.07 |
| Boat | 52631 | 39.06 | 52631 | 39.04 |
| Jet | 51020 | 41.31 | 51020 | 40.94 |
| Pepper | 51107 | 40.55 | 51107 | 40.61 |

Capacity in bytes and PSNR in dB and Cover images of size 512*512 .

## 6. CONCLUSIONS

In this paper, we have discussed a steganographic method for data hiding by using pixel value differencing which also guarantees that no pixel value will exceed the range 0 to 255 in stego-image. We have used original PVD method where pixel value does not cross the range, elsewhere proposed method has been used for embedding data. It gives same hiding capacity as the original PVD with acceptable stego-image quality.

## ACKNOWLEDGEMENTS

*The authors express gratitude to the Department of Computer Science and Engineering, University of Kalyani and the PURSE scheme of DST, Govt. of India, under which the research has been carried out.*

**Authors**

Dr. Jyotsna Kumar Mandal, M.Tech(Computer Science, University of Calcutta), Ph. D.(Engg. , Jadavpur University) in the field of Data Compression and Error Correction Techniques, Professor in Computer Science and Engineering, University of Kalyani, India. Life member of Computer Society of India since 1992 amd life member of Cryptology Research Society of India. Dean Faculty of Engineering, Technology and Management, working in the field of Network Security, Steganography, Remote Sensing & GIS Application, Image Processing. 25 years of teaching and research experiences. Eight scholars awarded Ph.D., one submitted and 8 scholars are pursuing. Total number of publications more than 228.

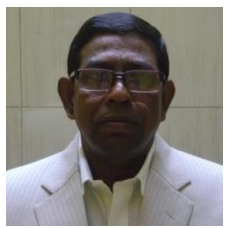

Debashis Das pursuing his M. Tech. in Computer Science and Engineering from University of Kalyani, under the Department of Computer Science and Engineering. 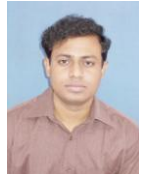